\documentclass{PoS}
\usepackage{dsfont}

\title{Simulating thimble regularization of lattice quantum field theories}

\ShortTitle{Simulating thimble regularization of lattice quantum field theories}

\author{\speaker{Francesco Di Renzo}\\
        University of Parma and INFN\\
        E-mail: \email{francesco.direnzo@unipr.it}}

\author{Giovanni Eruzzi\\
       University of Parma and INFN\\
       E-mail: \email{giovanni.eruzzi@pr.infn.it}}

\abstract{Monte Carlo simulations of lattice quantum field theories on Lefschetz thimbles are non trivial. We discuss a new Monte Carlo algorithm based on the idea of computing contributions to the functional integral which come from complete flow lines. The latter are the steepest ascent paths attached to critical points, i.e. the basic building blocks of thimbles. The measure to sample is thus dictated by the contribution of complete flow lines to the partition function. The algorithm is based on a heat bath sampling of the gaussian approximation of the thimble: this defines the proposals for a Metropolis-like accept/reject step. The effectiveness of the algorithm has been tested on a few models, e.g. the chiral random matrix model.
We also discuss thimble regularization of gauge theories, and in particular the successfull application to 0+1 dimensional QCD and the status and prospects for Yang-Mills theories.}

\FullConference{34th annual International Symposium on Lattice Field Theory\\
		24-30 July 2016\\
		University of Southampton, UK}

\begin{document}

\section{Introduction}
Following a seminal work by Witten \cite{Witten}, thimble
regularization of lattice field theories has been in recent years
introduced \cite{OurFirstTHMBL,Kikukawa}. In principle, it provides a
very clean solution to the so-called sign problem. All in all, it
amounts to deforming the original domain of integration of a given
field theory into a new one, which is made by one or more thimbles. 
Thimbles are manifolds which live in the complexification of the
original domain of integration. They are the union of the Steepest 
Ascent (SA) paths attached to critical points $\sigma$ of the (complexified)
action. Thimbles have the same real dimension of the original manifold
and on them the imaginary part of the action stays constant. The field
theoretic quantities one is interested in are expressed as
\begin{equation}
\label{eq:basicO}
\langle O \rangle = \frac{\sum_{\sigma} n_{\sigma} \,
  e^{-i\,S_I\left(p_{\sigma}\right)} \int_{\mathcal{J}_{\sigma}}
  \mathrm{d}y\, e^{-S_R} \,O\, e^{i\omega}}{\sum_{\sigma} n_{\sigma} \,
  e^{-i\,S_I\left(p_{\sigma}\right)} \int_{\mathcal{J}_{\sigma}}
  \mathrm{d}y\, e^{-S_R}  \, e^{i\omega}}
\end{equation}
where a positive measure $e^{-S_R}$ is in place and a constant
phase $e^{-i\,S_I\left(p_{\sigma}\right)}$ has been factored out of
the integral. In
the previous formula the denominator reconstructs the partition 
function $Z$. 
A so-called {\em residual phase} $e^{i \omega}$ is there that 
accounts for the relative orientation
between the canonical complex volume form and the real
volume form, characterizing the tangent space of the
thimble. \\
While the solution to the sign problem via a deformation of the 
integration domain is powerful and conceptually satisfying, thimbles 
are non-trivial manifolds, for which a local characterisation is
missing. In particular, devising Monte Carlo methods to sample
integrals on thimbles is a delicate issue.

\section{A new algorithm for thimble regularization}

A simple way to characterise points on a thimble goes through a
constructive approach. Given a critical point of the (complexified) 
action $S$, one can determine the tangent space to the thimble at that 
critical point. This is done by performing the Takagi factorization of
the Hessian of the action $S$ at the critical point: 
one is left with Takagi values ${\lambda_i>0}$ and Takagi vectors
$v^{(i)}$, which are a basis of the tangent space. The 
tangent space contains all the directions along which the SA
paths defined by\footnote{We denote generically by $z_i$ the (complex) 
degrees of freedom on the thimble. $t$ is the {\em time} coordinate
parametrizing the flow along the SA path.}
\begin{equation}
\frac{d}{dt}z_i = \frac{\partial \bar{S}}{\partial \bar{z}^i}
\end{equation}
leave the critical point. 
If we impose a normalization condition
\[
\sum_{i=1}^nn_i^2=\mathcal{R}
\]
all those directions are mapped to vectors 
\[
\sum_{i=1}^nn_iv^{(i)}.
\]
It is thus
quite natural to single out any given point on a thimble by the
correspondence
\begin{equation}
\label{eq:nNt}
\mathcal{J}_\sigma\ni z\leftrightarrow \left(\hat{n},t\right)\in S^{n-1}_{\mathcal{R}}\times\mathbb{R}
\end{equation}
with $S^{n-1}_{\mathcal{R}}$ the $(n-1)$-sphere of radius
$\sqrt{\mathcal{R}}$. In \cite{thimbleCRM} we made use of this
approach to solve a Chiral Random Matrix Model by means of thimble
regularization. It turns out that (\ref{eq:nNt}) amounts to computing a
Jacobian just like in the Faddeev-Popov approach to gauge fixing. 
For the sake of simplicity we restrict the discussion to cases in which the 
contribution attached to a single critical point $\sigma$ reconstructs
the entire functional integral (this happened to hold for the problem
discussed in \cite{thimbleCRM}, where however we discussed how
to proceed in a generic case). Let us define
\begin{equation} 
Z_\sigma=\int\limits_{\mathcal{J}_\sigma}\mathrm{d}^ny\,e^{-S_R}\label{eq:single_thimble_Z}
\end{equation}
With a slight abuse of terminology we will refer to this expression 
as a partition function. All in all, it can be rewritten
\begin{equation}
Z_\sigma=\int\mathcal{D}\hat{n}\,Z_{\hat{n}}\label{eq:totalZ_Zn}
\end{equation}
with the measure over $S^{n-1}_{\mathcal{R}}$
\[
\mathcal{D}\hat{n}\equiv\prod_{k=1}^n\mathrm{d} n_k\delta\left(\left|\vec{n}\right|^2-\mathcal{R}\right)
\]
and the \emph{partial} partition function
\begin{equation}
Z_{\hat{n}}=\int\limits_{-\infty}^{+\infty}\mathrm{d} t\,\Delta_{\hat{n}}(t)\,e^{-S_R(\hat{n},t)}\label{eq:partialZ}.
\end{equation}
The partition function has been decomposed in contributions
$Z_{\hat{n}}$ attached to SA paths ({\em aka} complete flow lines)
and $\Delta_{\hat{n}}(t)$ can be thought of as an extra contribution 
to the measure (on top of $e^{-S_R(\hat{n},t)}$) along the SA defined 
by the direction $\hat{n}$. The computation of $\Delta_{\hat{n}}(t)$
requires that one parallel transports the basis of
the tangent space at the critical point along the flow, to have a
basis $\{V^{(i)}\}$ at the (generic) point associated to the flow time $t$. More 
precisely, by assembling the $V^{(i)}$ into the matrix $V$, one finds
that 
\begin{equation} 
Z_{\hat{n}}=2\sum_{i=1}^n\lambda_in_i^2\int\limits_{-\infty}^{+\infty}\mathrm{d} t\,e^{-S_{\mathrm{eff}}(\hat{n},t)}\label{eq:partialZ_Seff}
\end{equation}
where the ${\lambda_i>0}$ are the Takagi values that were mentioned
above and the effective action $S_{\mathrm{eff}}$ is given by
\begin{equation} 
S_{\mathrm{eff}}(\hat{n},t)=S_R(\hat{n},t)-\log\left|\det V(t)\right|\label{eq:Zn_Seff}.
\end{equation}
In the (simplified) framework we are studying (a single contribution
to (\ref{eq:basicO}), coming from one thimble), it is easy to see that
the computation of (\ref{eq:basicO}) simply amounts to
\begin{equation}
\langle O\rangle=\frac{\langle O\,e^{i\,\omega}\rangle_\sigma}{\langle
  e^{i\,\omega}\rangle_\sigma}
\label{eq:obs_reweighted}
\end{equation}
where a reweighting with respect to the critical phase is in place and
we introduced the notation
\[
\langle\ldots\rangle_\sigma=\frac{\int\limits_{\mathcal{J}_\sigma}\mathrm{d}^ny\,\ldots\,e^{-S_R}}
{\int\limits_{\mathcal{J}_\sigma}\mathrm{d}^ny\,e^{-S_R}}.
\]
Making use of the representation (\ref{eq:nNt}), and thus of 
the same notation in which we wrote (\ref{eq:totalZ_Zn}) 
and (\ref{eq:partialZ}), one can now rephrase
\begin{equation}
\langle f \rangle_\sigma=
\frac{1}{Z_\sigma}\int\limits_{\mathcal{J}_\sigma}\mathrm{d}^ny\,f\,e^{-S_R}=\frac{1}{Z_\sigma}\int\mathrm{D}\hat{n}\;
(2\sum_{i=1}^n\lambda_in_i^2) \int\limits_{-\infty}^{+\infty}
\mathrm{d}t\,f(\hat{n},t)\,
e^{-S_{\mathrm{eff}}(\hat{n},t)}=\int\mathrm{D}\hat{n}\,\frac{Z_{\hat{n}}}{Z_\sigma}\,f_{\hat{n}}\label{eq:MC_expect_f}
\end{equation}
in which
\[
f_{\hat{n}}\equiv\frac{1}{Z_{\hat{n}}}
(2\sum_{i=1}^n\lambda_in_i^2) \int\limits_{-\infty}^{+\infty}
\mathrm{d}t\,f(\hat{n},t)\,
e^{-S_{\mathrm{eff}}(\hat{n},t)}
\]
almost looks like a functional integral along a single complete flow
line. (\ref{eq:MC_expect_f}) can be put at work in the computation of
(\ref{eq:obs_reweighted}) (with $f=O\,e^{i\,\omega}$ in the numerator 
and $f=e^{i\,\omega}$ in the denumerator). It is interesting to notice
that the $\frac{Z_{\hat{n}}}{Z_\sigma}$ factor contained in 
(\ref{eq:MC_expect_f}) is a legitimate (well normalized) weight, so 
that (\ref{eq:MC_expect_f}) is nothing but the average of the
$f_{\hat{n}}$, {\em i.e.} the average of the contributions that a 
given observable takes from complete flow lines. This average is 
computed in a given normalization, fixed by the
$\frac{Z_{\hat{n}}}{Z_\sigma}$ 
weight, which in turn represents the fraction of the
partition function which is provided by a single complete flow line.
In \cite{thimbleCRM} we made use of this approach for the computation
of (\ref{eq:obs_reweighted}), but we made no attempt at implementing
importance sampling with respect to the $\frac{Z_{\hat{n}}}{Z_\sigma}$
weight: computations were simply performed by flat, crude Monte Carlo,
{\em i.e.} extracting the directions $\hat{n}$ (flat) randomly.\\ 
Here we present a dynamic Monte Carlo, {\em i.e.} one 
performing importance sampling. 
This will amount to extract {\em directions} according to the probability 
$P(\hat{n})=\frac{Z_{\hat{n}}}{Z_\sigma}$. We proceed as follow. 
In our Markov chain we start from the current configuration ({\em
  i.e.} a direction $\hat{n}$) and we propose a new one 
({\em i.e.} a direction $\hat{n}'$) which is identical to $\hat{n}$ 
apart from two randomly chosen components, 
say $(n_i,n_j)$ with $i\neq j$. We define $C$ by
\[
C\equiv n_i^2+n_j^2=\mathcal{R}-\sum_{k\neq i,j}n_k^2
\]
which is fixed by the normalization
$\left|\vec{n}\right|=\sqrt{\mathcal{R}}$ and by the values of all
$\{n_k\}_{k\neq i,j}$. It exists a coordinate system in which we can 
now parametrize all the new values for $(n_i,n_j)$ by
\[
n_i=\sqrt{C}\,\cos\phi \;\;\;\;\; n_j=\sqrt{C}\,\sin\phi
\]
with $\phi\in[0,2\pi)$, and our aim is therefore to extract a value
for $\phi$. We now define a {\em gaussian} thimble: it is the  
thimble associated to a critical point of an action which only 
has quadratic fluctuations on top of the value at that critical point.
It is easy to find out that it is a flat thimble, and for it one can
compute
\begin{equation}
Z^{\mathcal{G}}_{\hat{n}}=2\sum_{i=1}^n\lambda_in_i^2\int\limits_{-\infty}^{+\infty}\mathrm{d} t\,e^{\;\;\sum\limits_{i=1}^n\lambda_i\,t-\frac{1}{2}\sum\limits_{i=1}^n\lambda_in_i^2e^{2\lambda_i t}}\label{eq:partialZ_gauss}.
\end{equation}
We now evaluate (\ref{eq:partialZ_gauss}) for $\hat{n}'$ as a 
function of $\phi$ and define the cumulative distribution function
\[
F^{\mathcal{G}}_{\hat{n}'}(\phi)\equiv\frac{\int\limits_0^{\phi}\mathrm{d}\varphi\,Z^{\mathcal{G}}_{\hat{n}'(\varphi)}}{\int\limits_0^{2\pi}\mathrm{d}\varphi\,Z^{\mathcal{G}}_{\hat{n}'(\varphi)}}.
\]
By extracting $\xi\in[0,1]$ uniformly distributed and computing
$\phi=F^{\mathcal{G}-1}_{\hat{n}'}(\xi)$\footnote{$F^{\mathcal{G}}_{\hat{n}'}(\phi)$,
  being the integral of a manifestly positive function, is
  monotonically increasing and can be easily inverted numerically.},
we obtain a $\phi$ ({\em i.e.} a $\hat{n}'$) that is distributed according to $P(\phi)\propto
Z_{\hat{n}'(\phi)}$. We now accept the proposed configuration with the
standard Metropolis test
\begin{equation}
P_{\mathrm{acc}}\left(\hat{n}'\bigr|\hat{n}\right)=\min\left\{1,\frac{Z_{\hat{n}'}}{Z_{\hat{n}}}
\frac{Z^{\mathcal{G}}_{\hat{n}}}{Z^{\mathcal{G}}_{\hat{n}'}}.
\right\}\label{eq:MC_flatmetroacc}
\end{equation}
This method turns out to be quite effective. In Figure 1 we present
results for the Chiral Random Matrix Model of \cite{thimbleCRM}.
For a given value of the mass parameter ($m=8$) we compare the results 
for the relevant condensate as computed from the crude Monte Carlo and
as computed from the new importance sampling method. Errorbars are
comparable despite the huge difference in the number of sampled
configurations. At a lower value of the mass parameter ($m=7$), it
turns out that crude Monte Carlo does not converge, while the new
method gets the correct result. 
\begin{figure}[ht] 
\centering
\includegraphics[scale=0.2]{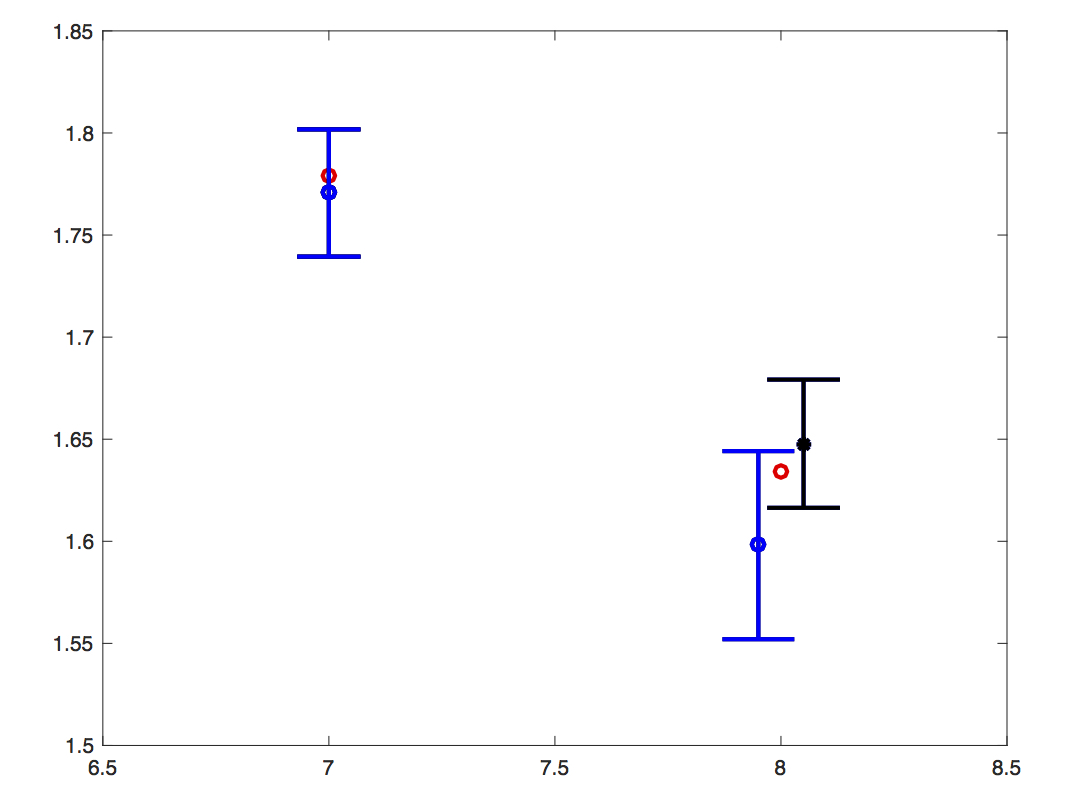}
\caption{Computation of an observable (chiral condensate) of a Chiral
  Random Matrix Theory via crude or dynamic Monte Carlo for different
  values of a mass parameter $m$. At $m=8$ the dynamic Monte Carlo
  obtains the result to the left, despite this comes from sampling
  $1300$ configurations, vs the $22000$ configurations sampled by
  crude Monte Carlo (result to the right). At $m=7$ crude Monte Carlo
  does not even converge, while dynamic gets the correct result.}
\label{fig:thimble_figure_stokes}
\end{figure}

\section{0+1 dim QCD}

We made use of the formulation induced by (\ref{eq:nNt}) also for 0+1
dimensional QCD \cite{QCD01}. As it is well known, this theory is in a convenient
gauge reduced to a single integral over $SU(3)$
\begin{equation}
Z_{N_f}=\int\limits_{\mathrm{SU}(3)}\mathrm{d} U\,{\det}^{N_f}\left(A\,\mathds{1}_{3\times 3}+e^{\mu/T}U+e^{-\mu/T}U^\dag\right)\label{eq:qcd01_Zint}
\end{equation}
The thimble machinery for $SU(N)$ was described in
\cite{lat2015gauge}, 
to which we refer the interested reader. Here the point we want to
make is that three critical points are there: does one need to take
into account all of them? This would contradict the hypothesis of the
main thimble dominance (notice that however this hypothesis is
supposed to hold at most in a thermodynamic limit we are quite apart
from in this system). The answer is illustrated in Figure 2. On the
left, we plot the (log of) the ratio between the partition functions 
(\ref{eq:single_thimble_Z}) computed on the thimbles attached to the 
non-trivial elements of $Z_3$ (they are equal) and the one computed 
on the thimble attached to the identity, {\em as computed in a
  semi-classical approximation}. The result is interesting. While the
complete analytic result does not know anything about thimbles, the 
semi-classical one attaches a different value to the computation in
the background of each critical
point. Results are for $N_f=2$ and $m=1$ and different values of the
ratio $\frac{\mu}{T}$. One can clearly notice that at certain values
of $\frac{\mu}{T}$ the contributions from thimbles other than the
identity are supposed to give a sizable contribution. On the right one
finds out that this is indeed the case. Different symbols in the
computation of the trace of Polyakov (second line) and anti-Polyakov
loop  (third line) refer to computations performed only keeping into
account the identity: they clearly miss the correct results. 

\begin{figure}[ht] 
\centering
\includegraphics[scale=0.23]{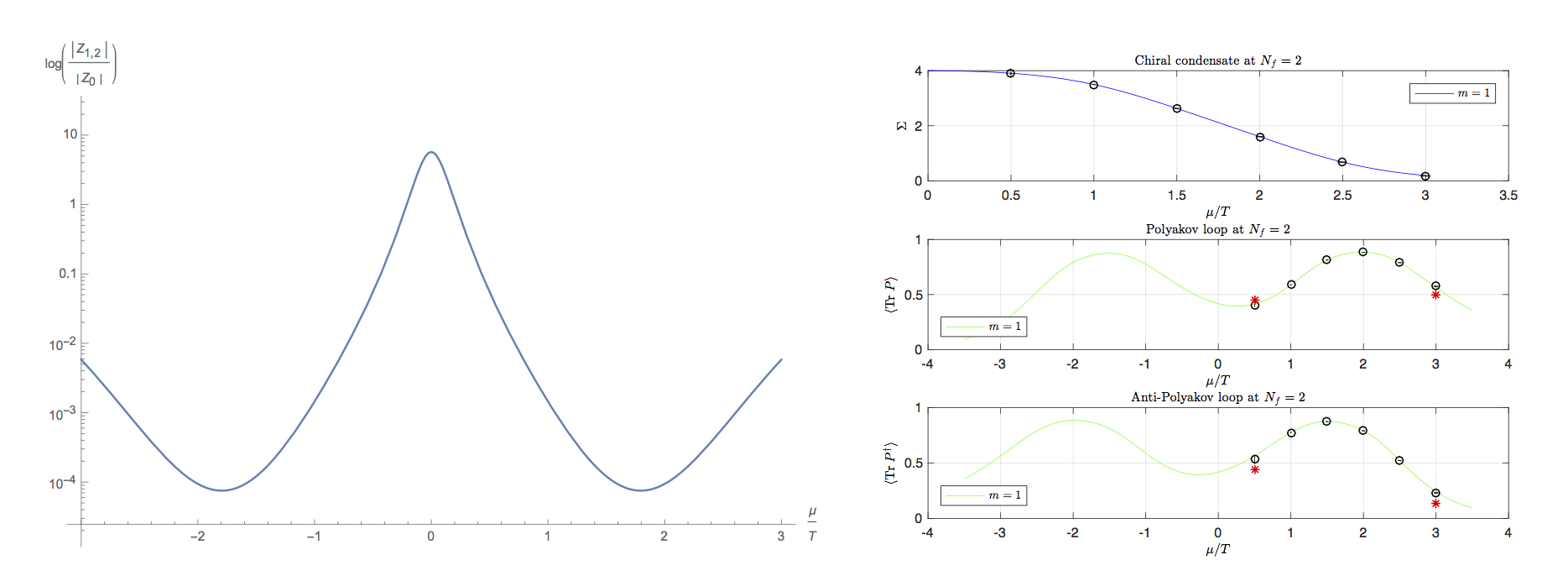}
\caption{Thimble computation of 0+1-dimensional QCD (see text).}
\label{fig:thimble_figure_stokes}
\end{figure}

\section{First steps in Yang-Mills theories}

We finally give a rough account of our first tests on Yang-Mills (YM)
theories. Again, the reader is referred to \cite{lat2015gauge} for
the relevant formalism. The theory at hand is the standard Yang-Mills
Wilson action, with a sign problem which is forced by plugging in a
complex value for the coupling $\beta$. Results can be tested versus analytic
ones in $d=2$. We tried to test in this framework the
viability of what we call the {\em gaussian approximation}. This could
be seen successfully at work in \cite{BoseTaming} (but see also comments
in \cite{thimbleCRM}). It amounts to performing a Langevin simulation on
the thimble. The drift term makes the system stay on the thimble by
very definition, and the problem is reduced to sample the noise term
on the tangent space. A solution was put forward in
\cite{OurFirstTHMBL}. In the gaussian approximation one simply
projects the noise over the tangent space at the critical point, thus
assuming the flat thimble to be a reasonable deformation of both the
original domain of integration and/or of the actual thimble. 
One does not obtain a constant imaginary part of the
action, but fluctuations are typically (even very) mild, at most
asking for a (viable) reweighting. Figure 3 displays gaussian
approximation results for the
action density of a $SU(2)$ YM theory in
$d=2$ on a (ridiculous...) $4^2$ lattice at $\beta=5e^{i 0.2}$. At
this value of the coupling semi-classical results would suggest the
gaussian approximation to work fairly well. This is indeed the
case. In particular numerical (gaussian approximation) results 
miss the analytic result without reweighting for the imaginary part
of the action (left), while a correct result is got once reweighting
is plugged in. All this is very preliminary. Still it is a very first
example of thimble regularization for gauge thoeries. 

\begin{figure}[ht] 
\centering
\includegraphics[scale=0.47]{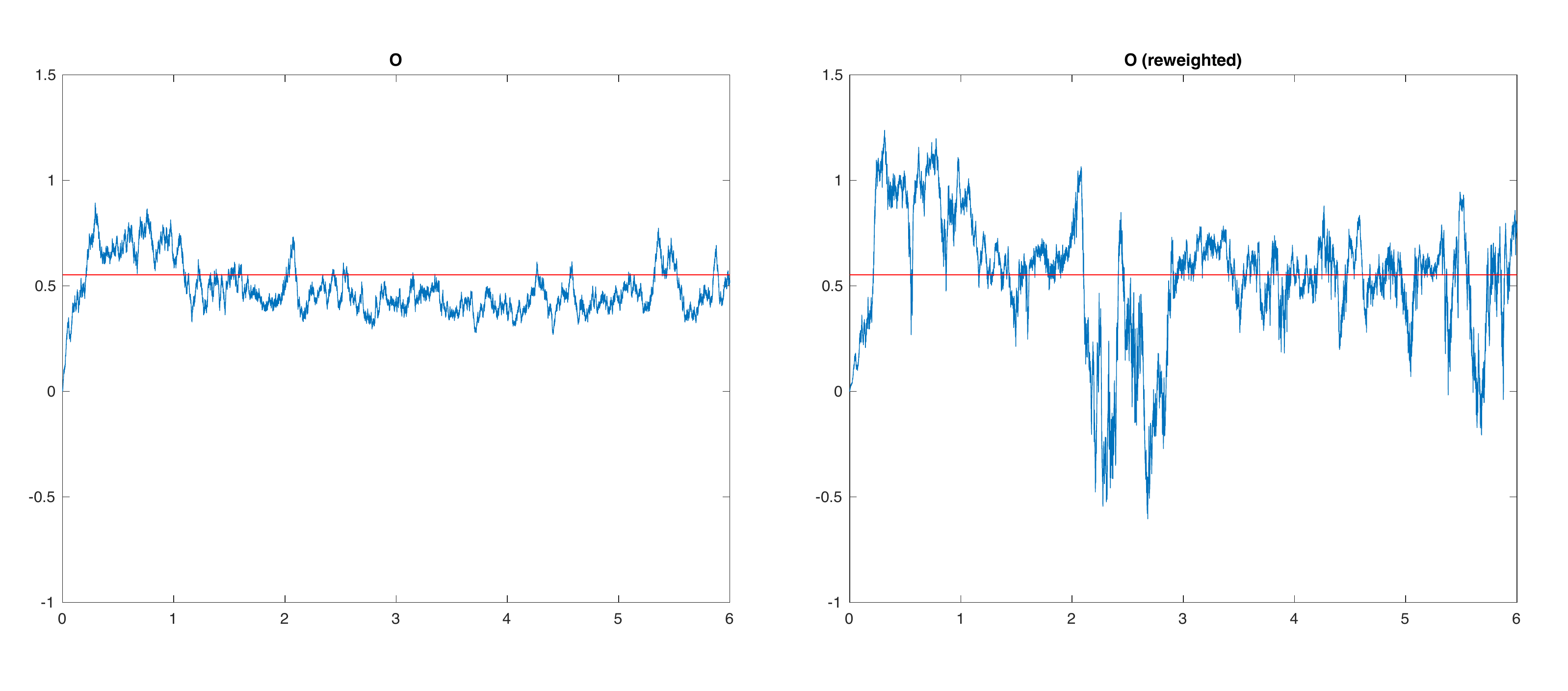}
\caption{Computation of the action density for $SU(2)$ YM theory in
$d=2$ on a $4^2$ lattice at $\beta=5e^{i 0.2}$. Red lines display the
analytic result. On the left there is
no reweighting for the imaginary part of the action, which is instead
there on the right. The correct result is got in the latter case,
despite quite sizable fluctuations (notice the different scales for $y$-axis).}
\label{fig:thimble_figure_stokes}
\end{figure}

% \section*{Conclusions}
% We reported on a new algorithm for simulating thimble regularization
% of lattice field theory, which was proven effective on a model (Chiral
% Random Matrix Model) which was previously computed by crude Monte
% Carlo. We also showed that first steps in thimble regularization of
% gauge thoeries are there: there is a huge amount of work to do, but
% some encouraging rsult is there.

\end{document}